\documentclass[a4paper,fleqn,usenatbib]{mnras}
\usepackage{newtxtext,newtxmath,hyperref}
\usepackage[T1]{fontenc}
\usepackage{ae,aecompl}
\usepackage[dvipdfmx]{graphicx}	
\usepackage{amsmath}	
\usepackage{amssymb}	
\usepackage{multirow,cases,colortbl}
\usepackage{url}
\usepackage{bm}
\usepackage{lscape}

\usepackage{ulem,color}

\newcommand{\MP}{\ensuremath{m_{\rm{p}}}}
\newcommand{\ME}{\ensuremath{m_{\rm{e}}}}
\newcommand{\SGM}{\ensuremath{\sigma_{\rm{T}}}}
\newcommand{\gr}{\ensuremath{\gamma}-ray }
\newcommand{\grs}{\ensuremath{\gamma}-rays }
\newcommand{\llgrb}{{\textit{ll}}GRB }
\newcommand{\llgrbs}{{\textit{ll}}GRBs }

\title[Generalized compactness limit]{Generalized compactness limit from an arbitrary viewing angle}

\author[Matsumoto,  Nakar \&  Piran]{Tatsuya Matsumoto,$^{1,2,3}$\thanks{E-mail: tatsuya.matsumoto@mail.huji.ac.il}
Ehud Nakar,$^{4}$
and Tsvi Piran$^{1}$
\\
$^{1}$Racah Institute of Physics, Hebrew University, Jerusalem, 91904, Israel\\
$^{2}$Department of Physics, Graduate School of Science, Kyoto University, Kyoto 606-8502, Japan\\
$^{3}$JSPS Research Fellow\\
$^{4}${The Raymond and Beverly Sackler School of Physics and Astronomy, Tel Aviv University, Tel Aviv 69978, Israel}
}
\pubyear{2018}

\begin{document}
\label{firstpage}
\pagerange{\pageref{firstpage}--\pageref{lastpage}}
\maketitle

\begin{abstract}
A  \gr source must have a limited optical depth to pair production. This simple condition, called compactness, implies that gamma-ray bursts (GRBs) must involve a highly relativistic  motion ($\Gamma \gtrsim 100$) giving the first and most important clue on their nature. So far, this condition 
has been discussed under the assumption that the \gr sources are viewed on-axis, that is, by an observer within the beaming cone of the relativistic source. Recently, following the detection of the weak short GRB 170817A, an extensive interest arose in the possibility that some \gr sources are viewed off-axis. We generalize  here the compactness formalism for an arbitrary viewing angle taking several possible opacity processes and \gr spectra into account. We find that for a given observables (peak luminosity, temporal variability, and spectra) the minimal Lorentz factor, $\Gamma_{\rm min}$, is obtained, as expected, for an on-axis observer.
More remarkably we find that compactness dictates also a maximal viewing angle, $\theta_{\rm max} \simeq 1/2\Gamma_{\rm min}$. Our limit implies for regular GRBs  a very small allowed viewing angle ($\lesssim10^{-2}\,\rm rad$), making it  extremely unlikely that they are viewed off-axis. For GRB 170817A  we confirm earlier results that rule out the possibility that the observed $\gamma$-rays were seen by an on-axis observer as a regular short GRB.
The short GRB 150101B was also suggested to be an off-axis event. We show that its maximal viewing angle $\lesssim0.05\,\rm rad$, which is inconsistent with the off-axis model.
Finally we show that for low luminosity GRBs, compactness does not exclude by itself an off-axis model, but when combined with other consideration this option is strongly disfavored.
\end{abstract}

\section{Introduction}
The first and most important clue on the nature of gamma-ray bursts (GRBs) arose from the compactness argument. 
The combination of huge luminosity, rapid variability, and a non-thermal spectrum implied that GRBs must involve a relativistic motion. A Newtonian calculation that uses the size implied by the temporal variability and the observed spectrum yields a huge optical depth ($\sim10^{13}$) for pair {production ($\gamma\gamma\to e^+e^-$)}. Such a source  would have had no photons above the electron's rest mass energy, $\ME c^2$, where $\ME$ and $c$ is the electron mass and the speed of light, respectively, and a thermal spectrum below it. This is the so-called compactness problem \citep{Ruderman1975,Schmidt1978}.
 It was  used as an argument in favor of the Galactic origin (as such bursts are of course much less energetic) or even for ``new physics''. Compactness problem is resolved, for bright cosmological GRB sources, by introducing a relativistic motion towards us (\citealt{Ruderman1975,Paczynski1986b,Goodman1986,Krolik&Pier1991}, {see also \citealt{Woltjer1966} in the context of compactness in active galactic nuclei).} The observed photon energy is boosted  and the observed timescale gets shorter. Both effects reduce the optical depth. 
For typical GRBs, the relativistic optical depth is $\sim 10^{13}/\Gamma^6$, where $\Gamma$ is the Lorentz factor, and one needs $\Gamma\gtrsim10^{2}$ to obtain the observed gamma-ray signal \citep{Piran1997,Piran1999,Lithwick&Sari2001}. This was the first and still strongest indication that GRBs involve a ultra-relativistic motion, as commonly understood today.

So far, the compactness problem has been discussed only for sources observed on-axis.
Here, we define ``on- (off-) axis'' view when an observer sees the source from the inside (outside)  the relativistic beaming cone whose angular width is  $\simeq1/\Gamma$. Clearly, off-axis observers see a \gr transient with a lower luminosity, lower typical photon energies and longer variability timescale than those observed by an on-axis one and therefore their compactness limits on the source are different.

There have been numerous suggestions that dimmer and softer transients, for example low-luminosity GRBs (\textit{ll}GRBs), are regular GRBs observed off-axis \citep[e.g.,][]{Nakamura1998,Eichler&Levinson1999,Woosley+1999,Ioka&Nakamura2001,Yamazaki+2003,Waxman2004}. Similar suggestions \citep{Goldstein+2017,Murguia-Berthier+2017b,Ioka&Nakamura2018} were made shortly following the detection of the very weak short GRB (sGRB) 170817A \citep{Abbott+2017e,Goldstein+2017,Savchenko+2017}. However, these suggestions did not take into account the compactness problem \citep{Kasliwal+2017}.  In an earlier paper \citep{Matsumoto+2019}, we have shown that compactness when combined with other conditions imposed by the afterglow observations \citep{Mooley+2018b} rules out the possibility that the observed \grs in sGRB 170817A arose from a regular sGRB jet viewed off-axis \citep[cf.][]{Eichler2017}.  We concluded that another \gr emission mechanism is needed, e.g. shock breakout of a cocoon \citep{Kasliwal+2017,Nakar+2018,Lazzati+2018,Kathirgamaraju+2018,Gottlieb+2018b,Bromberg+2018,Pozanenko+2018,Beloborodov+2018}.

Motivated by this example, we derive here the general compactness condition for an arbitrary \gr source for a general viewing angle. Our work generalizes \cite{Lithwick&Sari2001} that derived general compactness conditions for an on-axis observer.
In addition to the common condition that requires that the photons with the highest energy can escape from the source without pair {production}, we consider other processes that can dominate the \gr opacity:  the {Compton} opacity of pairs produced by annihilation of the high energy photons and the {Compton} opacity of the electrons in the outflow. At times, e.g. for sGRB 170817A, those latter conditions are more important than the first one that is more commonly considered.  To do that  we derive the optical depth to the different opacity sources as a function of the Lorentz factor and the viewing angle for the source.  

We organize this paper as follows:
In \S \ref{opacity sources} we review the processes that determines the \grs opacity at a source.
Next, we derive in \S \ref{general form of optical depth} an expression of the optical depth for given observables (peak luminosity, temporal variability, and spectra) as well as the source Lorentz factor and viewing angle.
Using this expression we obtain, in \S \ref{constraints},  limits on the Lorentz factor of the source  and on its viewing angle.
We apply the general argument in \S \ref{sec:application}  to typical GRBs, to the low-luminosity sGRBs 170817A and 150101B, and to several low-luminosity long GRBs (\llgrbs).
We summarize our results in \S \ref{summary and discussion}.

\section{Compactness}\label{compactness}
\subsection{Opacity sources of \grs}\label{opacity sources}
 We begin by recapitulating the essence of the  three  processes that determines the opacity at the source. 
The first two arise from pairs produced via {two-photon pair production} of the high energy photons. As such they depend on the observed spectrum. The third source of opacity arises from the electrons that exist within the source. While it does not depend on the spectrum its {significance} depends on whether the outflow is baryonic or not. 
Following \cite{Lithwick&Sari2001}, we denote these cases as  limits A, B, and C.

In the following we  relate the observables in the observer frame, denoted by $Q$ to those at the source rest frame, denoted by $Q^\prime$. The source is moving with a Lorentz factor $\Gamma$   ($\beta\equiv\sqrt{1-1/\Gamma^2}$) and the radiation field is roughly isotropic in 
the source local frame. The energy of the photons, $\epsilon$,  as seen in the different frames is related via the Doppler factor: 
\begin{align}
\delta_{\rm D}(\theta,\Gamma)\equiv\frac{1}{\Gamma(1-\beta\cos\theta)},
   \label{doppler factor}
\end{align}
such that 
\begin{align}
\epsilon=\delta_{\rm D}(\theta,\Gamma)\epsilon^\prime \ .
   \label{photon energy}
\end{align}
The angle $\theta $ is measured from the center of the gamma-ray emitting region (see below and Fig. \ref{fig picture}).\footnote{This definition of $\theta$ is different from the definition used by \cite{Matsumoto+2019}, where  $\Delta \theta_{\rm obs}$ is measured from the edge of the \gr emission region and not from its center.}
We ignore  redshift effects that can be trivially added later  (see \S \ref{constraints}).

\subsubsection{Limit A : Pair {production}}
The most common limit, denoted following \cite{Lithwick&Sari2001} limit A, arises by demanding that the highest energy 
photon observed with energy $\epsilon_{\rm max}$  escapes from the source without annihilating with other photons. A naive interpretation  is simply
$\epsilon_{\rm max}/\delta_{\rm D}(\theta,\Gamma) \le \ME c^2$. However this assumes that the spectrum has an unrealistic  sharp cut-off at $\epsilon_{\rm max}$ (e.g., in typical GRBs the number of photons above $\epsilon_{\rm max}$ should drop by $\sim 10$ orders of magnitude  before the source becomes optically thin to $\epsilon_{\rm max}$ photons under this assumption). 
There is no reason to assume such a coincidence since   $\epsilon_{\rm max}$ 
depends on the detector. Either the flux above $\epsilon_{\rm max}$ is too low to be detected or it is just the upper limit of the detector's window. 
A priori it is not reasonable to assume that the spectrum breaks just at this energy.  It is most natural to extrapolate the observed spectrum  to higher energies. 
The assumption that the spectrum can be extrapolated upwards is reasonable in the case that the upper part of the spectrum is a power-law and can be considered as conservative when considering an exponential cutoff. Note that by extrapolating one assumes only that photons more energetic than $\epsilon_{\rm max}$ are present at the source, not that they escape from it (namely the source does not have to be optically thin to photons with $\epsilon>\epsilon_{\rm max}$).
To estimate the optical depth for pair {production} of a photon with $\epsilon_{\rm max}$ we define 
 $\epsilon_{\rm th,A}(\epsilon_{\rm max})$ the minimal energy of the photons with which an 
 $\epsilon_{\rm max}$ photon can annihilate:
\begin{align} 
\epsilon_{\rm th,A}(\epsilon_{\rm max})=\frac{\big[\delta_{\rm D}(\theta,\Gamma)\ME{c^2}\big]^2}{\epsilon_{\rm max}}\,;\,\,\,\text{limit A} \ .
   \label{energy limit A}
\end{align}
We then  calculate  the fraction, $f$,  of photons above $\epsilon_{\rm th,A}$ and demand that their optical depth is sufficiently small. 
Limit A has been commonly used in the GRB literatures but, depending in particular on the spectral shape, it is not always the most stringent one.

\subsubsection{Limit B : Scattering  by  pairs}
For limit B, we consider the opacity {attributable to} the pairs created by self-annihilating photons.
The energy threshold of photons which can self-annihilate is given by
\begin{align} 
\epsilon_{\rm th,B}=\delta_{\rm D}(\theta,\Gamma)\ME{c^2}\,;\,\,\,\text{limit B}.
   \label{energy limit B}
\end{align}
We approximate the number of pairs as the number of photons with energy larger than the threshold energy, and calculate the fraction $f$, of photons above this threshold for a given observed spectrum.

Limit B is less known than limit A. However, at times it is more important. It dominates for bursts with a small maximal observed photon energy $\epsilon_{\rm max}$, which is in particular the case when the bursts are dim or when the upper end of the observed spectrum is exponential (as in the case of a Comptonized power-law) rather than a power-law. Note that also here, like in limit A, the limit is dictated by an extrapolation of the observed spectrum to energies above the highest observed photon energy, and when a Comptonized spectrum is considered then this extrapolation is the most natural one.

\subsubsection{Limit C : Electron scattering}
A third source of opacity  is  optical depth  of the source's electrons  that accompany the baryons in the outflow.
The total electron number can be estimated,  from the baryon number, which in turn is determined by the energy budget. For any baryonic outflow, the  thermal energy  in the rest frame is larger than the total rest frame {emitted} photon energy.
The thermal energy per baryon is given by $w-\MP c^2$, where $w$ and $\MP$ are the enthalpy per baryon and the proton mass, respectively.
An extreme assumption for a baryonic outflow is that it was accelerated to its Lorentz factor $\Gamma$ from rest and crossed a shock just before the emission took place. In that case $w \simeq \Gamma \MP c^2$. {Note that \cite{Lithwick&Sari2001} assume here that the available baryonic energy is $w-\MP c^2\simeq \MP c^2$. Our estimate is larger by a factor of $\Gamma-1$ and our limit is, correspondingly, less restrictive. }
The energetics argument limits the total baryon number as
\begin{align}
\Gamma\MP c^2N_{\rm baryon}^\prime\gtrsim \epsilon_{\rm p}^\prime N_{\rm \gamma}^\prime,
   \label{baryon energetics}
\end{align}
where {$\epsilon_{\rm p}^\prime$ and $N_{\gamma}^\prime$ are the observed peak photon energy (peak energy of the energy flux)} and the total photon number in the emitting region, respectively, such that the right-hand-side is approximately the total rest frame radiation energy.
Clearly this estimation is valid only in the case that the energy source is the  energy carried by  the baryons (e.g. for an internal-shock model but not a Poynting flux dominated outflow).

For normal GRBs, limit C gives the weakest limit, hence not much attention was paid to it in terms of limit on the Lorentz factor. Instead it was used to put a lower limit on the emission radius. 
In this limit, pair {production} has nothing to do with the opacity, and  the optical depth depends only on total observed radiation energy and it is independent of the specific observed  spectrum.

\subsection{The optical depth}\label{general form of optical depth}
We consider a general \gr transient,  characterized by the following observables: (a) an isotropic \gr luminosity $L_{\rm \gamma,iso}$, (b) a variable timescale (duration of an observed single pulse) $\delta t$, and (c) the photon spectrum $dN/d\epsilon$ with a peak energy $\epsilon_{\rm p}$ (see Table \ref{table summary}). 
Using these observables, we evaluate the optical depth of the \gr transient taking the different limits into account.
Note  that in this section we do not take  redshift effects into account. However those are inserted later  when we estimate the final results shown in Table \ref{table summary2}.

The optical depth of \grs in the rest frame is given by \citep[e.g.,][]{Nakar2007}
\begin{align}
\tau\simeq\frac{\sigma N_{\rm s}^\prime l^\prime}{\pi \theta_\gamma^2R^2\Delta R^\prime}.
   \label{tau_org}
\end{align}
$\sigma$ is the relevant cross section, $l^\prime$ is the path length of the photons within the region before escaping to the observer, and  $N_{\rm s}^\prime$ is the number of annihilating photons in limit A, number of produced pairs in limit B, and number of source electrons in limit C. 
$\theta_\gamma$, $R$, and $\Delta R^\prime$ are the characteristic angle of the \gr emitting region, radial distance where the emission episode takes place, and the {thickness} of the \gr emitting region, respectively.
The angle and distance are measured in the lab frame.

{The cross section for pair production peaks slightly above the  threshold (given by Eq. \ref{energy limit A}) and then it decreases quickly with energy.}
For a power-law photon spectrum with a photon index  $-2$, the average value of the cross section becomes $(11/180)\sigma_{\rm T}$ where  $\SGM$  is the Thomson cross section,  \citep{Svensson1987,Lithwick&Sari2001}.
We use this factor  for a power-law spectrum (PL). We don't  use it for a cut-off power-law spectrum  that declines sharply at high energy.
In the following, we still use the notation $\SGM$ for simplicity, but   restore the coefficient in the summarizing Table \ref{table summary2}.

\begin{figure}
\begin{center}
\includegraphics[width=85mm]{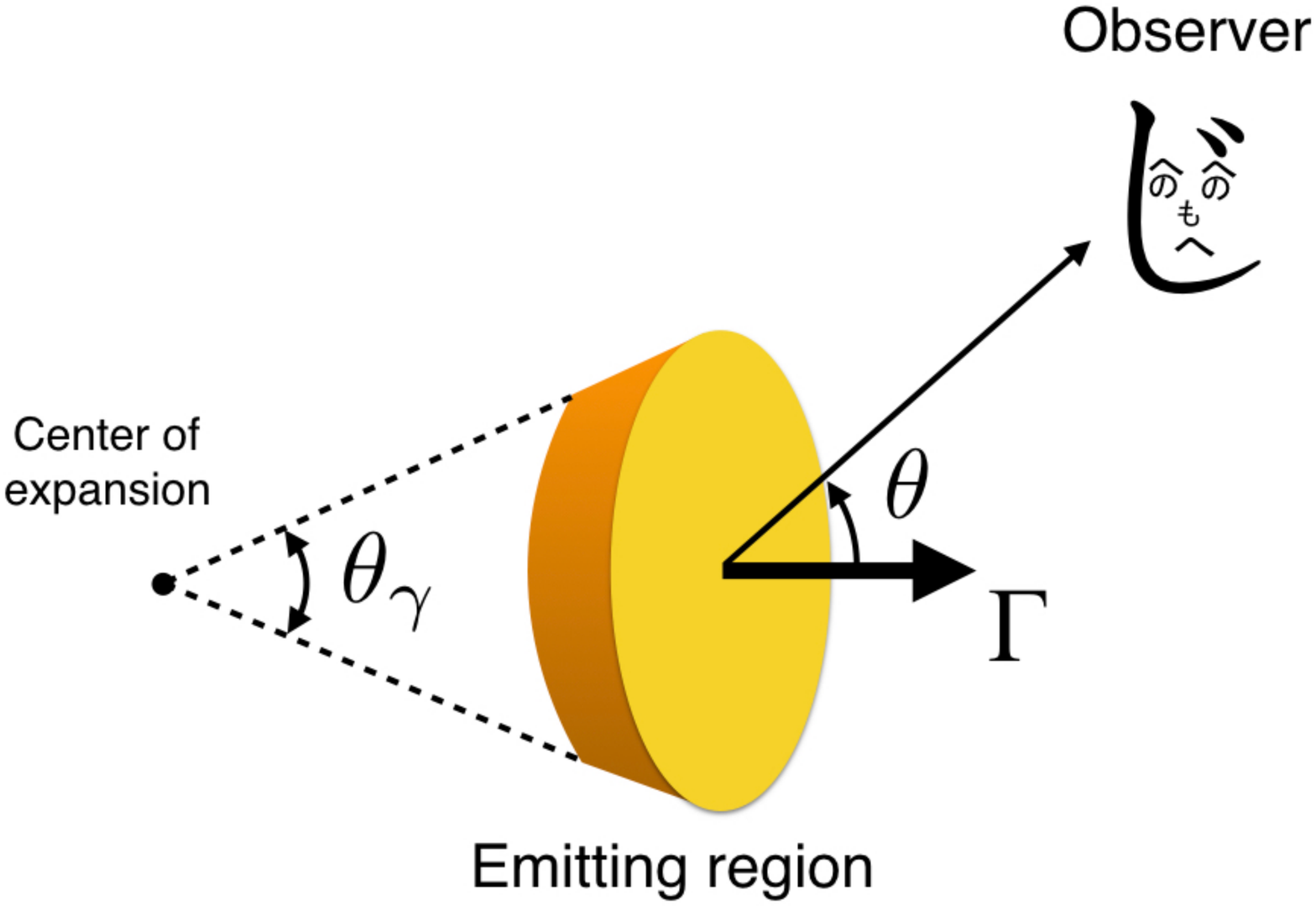}
\caption{Schematic picture of a \gr emitting region and an observer. Over this region, the luminosity and Lorentz factor are uniform.}
\label{fig picture}
\end{center}
\end{figure}

Generally, the outflow that produces the  \gr emission has an angular structure, e.g., luminosity and Lorentz factor distributions that depend on the angle.
Each point on the outflow contributes to the observed \gr flux with different intensity and Doppler boost.
In \cite{Matsumoto+2019}, we showed that, because of the sensitive dependence of the Doppler boost on the viewing angle, unless there is an extreme fine tuning a small patch of the outflow dominates the observed \grs.
Thus, we can reasonably assume that the luminosity and Lorentz factor over the patch are approximately constant.
We denote this region as the \gr emitting region and approximate the geometry of the emitting source 
 by a characteristic angular size $\theta_\gamma$ 
over which the luminosity and the Lorentz factor do not vary significantly.  It is viewed from an angle $\theta$, measured from the center of this region (see Fig. \ref{fig picture}).

For a small viewing angle and a large Lorentz factor, the angular variation of the Doppler factor  depends on the product $\Gamma \theta$.  Hence, for $\theta<1/\Gamma$, the observer is on-axis and relativistic beaming dictates the size of the \gr emitting region: $\theta_\gamma \le 1/\Gamma$. 
For $\theta>1/\Gamma$, the observer is off-axis and the observed \gr emission is suppressed by the Doppler de-beaming.   Hence, only a region of size $\theta$ can contribute significantly. 
Overall, the size of the \gr emitting region satisfies:
\begin{align}
\theta_{\gamma}\leq{\rm max}\biggl\{\frac{1}{\Gamma},\theta\biggl\}.
\end{align}

We turn now to evaluate the various quantities that determine the optical depth (Eq. \ref{tau_org}) using the observables.
For limits A and B, we write $N_{\rm s}^\prime$  as
\begin{align}
N_{\rm s}^\prime=fN_{\gamma}^\prime \ . 
   \label{scatterer number}
\end{align}
Here,  $f$ is the fraction of photons whose energy is larger than the thresholds $\epsilon_{\rm th,A}$ 
(Eq. \ref{energy limit A})  or $\epsilon_{\rm th,B}$  (Eq. \ref{energy limit B}).
$N_{\rm \gamma}$ is the total number of photons:
\begin{align}
N_{\rm \gamma}^\prime
\simeq\frac{4\pi{d^2}S_\gamma
}{\delta_{\rm D}^2(\theta,\Gamma)\epsilon_{\rm p}} 
\simeq\frac{ L_{\rm \gamma,iso} \delta t 
}{\delta_{\rm D}^2(\theta,\Gamma)\epsilon_{\rm p}} ,
   \label{photon number}
\end{align}
where $d$ is  distance to the burst , $S_\gamma \simeq  L_{\rm \gamma,iso} \delta t /4\pi d^2$  is  observed \gr fluence 
during $\delta t$ and  $L_{\rm \gamma,iso} $ the isotropic equivalent luminosity. In {deriving} this relation, we have used the transformation of the solid angles $\Delta\Omega=\Delta\Omega^\prime/\delta_{\rm D}^2(\theta,\Gamma)$, and assumed that the source radiates photons isotropically in the rest frame.

For a normalized photon spectrum, the photon fraction, $f$,  is calculated as
\begin{align}
f=\int_{\epsilon_{\rm th}}^\infty\frac{dN}{d\epsilon}d\epsilon \ , 
\end{align}
where $\epsilon_{\rm th}=\epsilon_{\rm th,A}$ or $\epsilon_{\rm th,B}$ and we assume that the observed high end of the photon spectrum extends to higher energies. 
We consider two spectra: a power-law in the relevant (high energy) segment (denoted PL)
and  a power-law with an exponential cut-off, also called a Comptonized spectrum (CPL):
\begin{subnumcases}
{\frac{dN}{d\epsilon}\propto}
\epsilon^{\beta_p}&\text{; PL,}
   \label{spectrum pl}\\
\epsilon^{\alpha_p}\exp\biggl[-\frac{\epsilon(\alpha_p+2)}{\epsilon_{\rm p}}\biggl]&\text{; CPL,}
   \label{spectrum cut-off pl}
 \end{subnumcases}
where   (following here the standard notation in the GRB literature) $\beta_p$  and $\alpha_p$  are  the spectral indices for the high energy part of the Band function in the PL case and the power-law of the CPL case, respectively.
The fraction of annihilating photons is given by:
\begin{subnumcases}
{f\sim}
\biggl(\frac{\epsilon_{\rm th}}{\epsilon_{\rm p}}\biggl)^{\beta_p+1}&\text{; PL,}
   \label{photon fraction PL}\\
\biggl(\frac{\epsilon_{\rm th}(\alpha_p+2)}{\epsilon_{\rm p}}\biggl)^{\alpha_p}\exp\biggl(-\frac{\epsilon_{\rm th}(\alpha_p+2)}{\epsilon_{\rm p}}\biggl)&\text{; CPL.}
   \label{photon fraction CPL}
\end{subnumcases}
In the PL case, we consider only the high energy part of the  spectrum and  we normalize the spectrum at the peak energy $\epsilon_{\rm p}$.\footnote{Strictly speaking, for the Band function \citep{Band+1993}, the peak energy should be replaced with $\epsilon_{\rm p}/(\alpha_{p,\rm Band}+2)$ where $\alpha_{p,\rm Band}$ is the low energy power-law index of the Band function.}

In limit C, electrons associated with baryons in the outflow scatter  the photons.
The total baryon number is given by Eq. \eqref{baryon energetics} :
\begin{align}
N_{\rm s}^\prime=N_{\rm baryon}^\prime\gtrsim\frac{\epsilon_{\rm p}}{\Gamma\delta_{\rm D}(\theta,\Gamma)\MP c^2}N_{\gamma}^\prime \ .
\end{align}
Since this value is also proportional to $N_{\gamma}^\prime$, we can use the same form of  $N_{\rm s}^\prime$ as in  limits A and B by defining the fraction, $f$,  as
\begin{align}
f\equiv\frac{\epsilon_{\rm p}}{\Gamma\delta_{\rm D}(\theta,\Gamma)m_{\rm p}c^2}\,;\,\,\,\text{limit C} \ .
   \label{photon fraction limit C} 
\end{align}

The crux of the argument in the off-axis case is to estimate, using the observed variability time scale $\delta t$,  the emission radius, $R$, and use it to  obtain a limit on the optical depth $\tau$. 
The observed pulse duration is the sum of three times.
The angular spreading time
\begin{align}
\delta t_{\rm ang}
=\frac{2R}{c}\begin{cases}
\sin(\theta)\sin\Big(\frac{\theta_\gamma}{2}\Big)&;\theta>\frac{\theta_\gamma}{2}\ ,\\
1- \cos\Big[\frac{1}{2}\Big(\theta+\frac{\theta_\gamma}{2}\Big)\Big]&;\theta<\frac{\theta_\gamma}{2} \  , 
\end{cases}
   \label{angular spreading time}
\end{align}
the radial timescale
\begin{align}
\delta t_{\rm radial}&\equiv\frac{dR}{c\beta}\Big[1-\beta\cos\theta\Big]=\frac{dR}{c\beta\Gamma\delta_{\rm D}(\theta,\Gamma)} \ , 
   \label{radial time}
\end{align}
where $dR \leq R$ is the distance crossed by the source while it was active, and the light crossing time $\delta t_{\rm lc}$ of the photon shell. Clearly $\delta t \simeq
\max\{\delta t_{\rm ang},\delta t_{\rm radial},\delta t_{\rm lc}\}$. 

In Appendix A, we show that if $\delta t_{\rm lc} >(\delta t_{\rm radial},\delta t_{\rm ang})$, i.e. the light crossing time is the longest, then the limits on the opacity are more constraining than those obtained in case that it is not. Therefore, we consider conservatively  $\delta t_{\rm lc} < \delta t_{\rm radial}$, in which case $l^\prime \simeq \Delta R^\prime$ as a photon emitted from one side of the emitting region towards the observer crosses the entire emitting region by the time that the radius doubles  (this point is also shown in appendix A). Taking $l^\prime \simeq \Delta R^\prime$, Eq. \eqref{tau_org} becomes: 
\begin{align}
\tau\simeq \frac{\sigma N_{\rm s}^\prime }{\pi \theta_\gamma^2R^2 } \ .
   \label{tau}
\end{align}
Next we examine the limits on $\tau$ when $R$ is constrained by the condition $\delta t = \max\{\delta t_{\rm ang},\delta t_{\rm radial}\}$. We show in appendix B that in the off-axis case the limit obtained under the assumption that $\delta t=\delta t_{\rm ang}$, is similar to the one obtained in case that $\delta t=\delta t_{\rm rad}$ and $dR=R$. We also show that this is also the case for an on-axis observer, unless there is an anomalously narrow source with $\theta_\gamma \ll 1/\Gamma$. We therefore continue deriving the limit on $\tau$ by by setting $\delta t=\delta t_{\rm rad}$ where $dR=R$, but we stress that the limit we obtain is applicable for any type of source when observed off-axis, including a point source that radiates instantaneously in time, and to any source with $\theta_\gamma \sim 1/\Gamma$ when observed on-axis.

Substituting Eqs. \eqref{scatterer number}, \eqref{photon number},  and \eqref{radial time} with $dR=R$ into Eq. \eqref{tau} we
find that $\theta_\gamma$ appears in the denominator and hence $\theta_\gamma$  has to attain its maximal value, $\theta_\gamma={\rm max}\{1/\Gamma,\theta\}$, 
to minimize $\tau$.  
Overall we obtain a lower limit on the optical depth: 
 \begin{align}
\tau \geq \tau_{\rm  min}&= \biggl \{\frac{\sigma_{\rm T}}{\pi c^2} \frac{ L_{\rm \gamma,iso}}{\epsilon_{\rm p}\delta t
} \biggl \}\biggl [ \frac{f}{[{\max}\{1/\Gamma,\theta\}]^2\beta\Gamma^2\delta_{\rm D}^4(\theta,\Gamma)} \biggl ] \nonumber \\
&= {\cal L} \biggl [ \frac{f}{[{\rm max}\{1/\Gamma,\theta\}]^2\beta\Gamma^2\delta_{\rm D}^4(\theta,\Gamma)} \biggl ] \ ,
   \label{tau general}
\end{align}
where we define the dimensionless observable
\begin{align}
{\cal L}&\equiv \frac{\sigma_{\rm T}}{\pi c^2}\frac{L_{\rm \gamma,iso}}{\epsilon_{\rm p}\delta t}
   \label{L} \nonumber \\
&\simeq1.5\times10^{13}\,\Big(\frac{ L_{\rm \gamma,iso}}{10^{51}\, {\rm ergs/s}}\Big)\Big(\frac{\epsilon_{\rm p}}{100\,{\rm keV}}\Big)^{-1} \Big( \frac{\delta t
}{0.1\, {\rm s}}\Big)^{-1} \ . 
\end{align}
Note that ${\cal L}$ should be multiplied by the factor $11/180$  for limit A and  a PL spectrum (see Table \ref{table summary2}).

Eq. \eqref{tau general}  together with an estimate of $f$ (obtained using  Eq. \ref{photon fraction PL} or \ref{photon fraction CPL} with energy thresholds Eq. \ref{energy limit A} or \ref{energy limit B} for limits A and B, and Eq. \ref{photon fraction limit C} for limit C) is a general form of the minimal optical depth for an observer from an arbitrary viewing angle.  We stress that this is only a lower limit on the opacity obtained under optimal conditions that minimize the opacity. The actual conditions could be more stringent leading to a stronger limits on the Lorentz factor and on the viewing angle. 
Eq. \eqref{tau general} generalizes the common expression for the optical depth for an  observer at $\theta = 0$  obtained by \cite{Lithwick&Sari2001}. As we see later this $\theta = 0$ expression is a very good approximation for all on-axis observers. 

Table \ref{table summary2} summarizes the different possible forms  that the expression (Eq. \ref{tau general}) takes for the various cases discussed here. We mark in  colors the cases that correspond to typical GRBs and \llgrbs {which are the subclass of long GRBs with low luminosity (see \S \ref{grb980425}, for more detailed properties of \llgrbs).}

\begin{table*}
\begin{center}
\caption{Summary of  observables.}
\label{table summary}
\begin{tabular}{llrr}
\hline\hline
\multicolumn{2}{c}{Observables}&\multicolumn{2}{c}{Typical values used in Fig. \ref{fig sample}}\\
Symbol&Definition&left&right\\
\hline
$L_{\rm \gamma,iso}$&Isotropic \gr luminosity&$10^{47}{\rm erg\, s^{-1}}$&$10^{53}\,\rm erg\,s^{-1}$\\
$\delta t
$&Variable timescale&$0.1\,\rm s$&$0.01\,\rm s$\\
$\epsilon_{\rm p}$&Peak energy of photons&$0.5\,\rm MeV$&$4\,\rm MeV$\\
$\epsilon_{\rm max}$&Maximum energy of photons&$0.5\,\rm MeV$&$30\,\rm GeV$\\
$\beta_p$&Spectral index (PL)&$-2$&$-2$\\
$\alpha_p$&Spectral index (CPL)&$-1$&$-1$\\
$z$&redshift&0&1\\
\hline
\end{tabular}
\end{center}
\end{table*}

\begin{table*}
\begin{center}
\caption{Summary of lower limits on the Doppler factor $\delta_{\rm D}$, the minimal Lorentz factor $\Gamma_{\rm min}$, and the maximal viewing angle $\theta_{\rm max}$. The colored cells mark the relevant limits for typical GRBs (yellow) and \llgrbs {(pink, see \S \ref{grb980425} for the properties of \llgrbs).}}
\label{table summary2}
\begin{tabular}{cc||cccccc}
\hline\hline
Spectrum
&Limit&${\cal L}$&${\cal E}$&$\tau_{\rm min}$&$\delta_{\rm D,min}$&$\Gamma_{\rm min}$&$\theta_{\rm max}$\\
\hline
\multirow{4}{*}{PL}&limit A&\cellcolor[rgb]{1,1,0.5}$\frac{(11/180)\sigma_{\rm T}}{\pi c^2}\frac{L_{\rm \gamma,iso}}{\epsilon_{\rm p}\delta t
}$&\cellcolor[rgb]{1,1,0.5}$\frac{\epsilon_{\rm p}\epsilon_{\rm max}(1+z)^2}{(m_{\rm e}c^2)^2}$&\cellcolor[rgb]{1,1,0.5}$\frac{{\cal L}}{{\cal E}^{\beta_p+1}}\delta_{\rm D}^{2(\beta_p-1)}$&\cellcolor[rgb]{1,1,0.5}$\big(\frac{{\cal E}^{\beta_p+1}}{{\cal L}}\big)^{\frac{1}{2(\beta_p-1)}}$&\cellcolor[rgb]{1,1,0.5}$\frac{1}{2}\big(\frac{{\cal E}^{\beta_p+1}}{{\cal L}}\big)^{\frac{1}{2(\beta_p-1)}}$&\cellcolor[rgb]{1,1,0.5}$\frac{1}{2}\big(\frac{{\cal E}^{\beta_p+1}}{{\cal L}}\big)^{\frac{1}{2(1-\beta_p)}}$\\
&limit B&\multirow{2}{*}{$\frac{\sigma_{\rm T}}{\pi c^2}\frac{L_{\rm \gamma,iso}}{\epsilon_{\rm p}\delta t
}$}&$\frac{\epsilon_{\rm p}(1+z)}{m_{\rm e}c^2}$&$\frac{{\cal L}}{{\cal E}^{\beta_p+1}}\delta_{\rm D}^{\beta_p-3}$&$\big(\frac{{\cal E}^{\beta_p+1}}{{\cal L}}\big)^{\frac{1}{\beta_p-3}}$&$\frac{1}{2}\big(\frac{{\cal E}^{\beta_p+1}}{{\cal L}}\big)^{\frac{1}{\beta_p-3}}$&$\frac{1}{2}\big(\frac{{\cal E}^{\beta_p+1}}{{\cal L}}\big)^{\frac{1}{3-\beta_p}}$\\
&limit C&&$\frac{\epsilon_{\rm p}(1+z)}{\MP c^2}$&$\frac{{\cal LE}}{\Gamma\delta_{\rm D}^5}$&-&$\big(\frac{{\cal LE}}{32}\big)^{1/6}$&$\big(2{\cal LE}\big)^{-1/6}$\\
\hline
\multirow{4}{*}{CPL}&limit A&&$\frac{\epsilon_{\rm p}\epsilon_{\rm max}(1+z)^2}{(m_{\rm e}c^2)^2(\alpha_p+2)}$&$\frac{{\cal L}}{{\cal E}^{\alpha_p}}\delta_{\rm D}^{2(\alpha_p-2)}\exp\big(-\frac{\delta_{\rm D}^2}{\cal E}\big)$&Eq. \eqref{tau parameterized}*&$\frac{\delta_{\rm D,min}}{2}$&$\frac{1}{2\delta_{\rm D,min}}$\\
&limit B&\cellcolor[rgb]{1,0.7,0.7}$\frac{\sigma_{\rm T}}{\pi c^2}\frac{L_{\rm \gamma,iso}}{\epsilon_{\rm p}\delta t
}$&\cellcolor[rgb]{1,0.7,0.7}$\frac{\epsilon_{\rm p}(1+z)}{m_{\rm e}c^2(\alpha_p+2)}$&\cellcolor[rgb]{1,0.7,0.7}$\frac{{\cal L}}{{\cal E}^{\alpha_p}}\delta_{\rm D}^{\alpha_p-4}\exp\big(-\frac{\delta_{\rm D}}{\cal E}\big)$&\cellcolor[rgb]{1,0.7,0.7}
Eq. \eqref{tau parameterized}*&\cellcolor[rgb]{1,0.7,0.7}$\frac{\delta_{\rm D,min}}{2}$&\cellcolor[rgb]{1,0.7,0.7}$\frac{1}{2\delta_{\rm D,min}}$\\
&limit C&\cellcolor[rgb]{1,0.7,0.7}&\cellcolor[rgb]{1,0.7,0.7}$\frac{\epsilon_{\rm p}(1+z)}{\MP c^2}$&\cellcolor[rgb]{1,0.7,0.7}$\frac{{\cal LE}}{\Gamma\delta_{\rm D}^5}$&\cellcolor[rgb]{1,0.7,0.7}-&\cellcolor[rgb]{1,0.7,0.7}$\big(\frac{{\cal LE}}{32}\big)^{1/6}$&\cellcolor[rgb]{1,0.7,0.7}$\big(2{\cal LE}\big)^{-1/6}$\\
\hline
\multicolumn{8}{l}{Limit A: $\epsilon_{\rm max}$ photon can escape from the source without pair {production}.}\\
\multicolumn{8}{l}{Limit B: the emitting region is optically thin to Thomson scattering by pairs created by the \gr photons .}\\
\multicolumn{8}{l}{Limit C:  the emitting region is optically thin to scattering by electrons in the source. The limit is relevant for a baryonic outflow and}\\
\multicolumn{8}{l}{independent of the spectral shape.}\\
\multicolumn{8}{l}{* There is no analytic solution.}\\
\end{tabular}
\end{center}
\end{table*}

\subsection{Limits on $\Gamma$ and $\theta$}\label{constraints}
The fact that we observe bright \grs implies that the optical depth of the source is limited. The reasons for this limit depend on the spectrum. If the spectrum is highly non-thermal, as in the case of a high-energy power-law, then $\tau \lesssim 1$ since $\tau \gg 1$ leads to either a blackbody or a Wien spectrum. This argument is not applicable to a Comptonized spectrum since it can be a result of a sum of blackbody  or Wein spectra with different temperatures (the sum is needed to produce a shallow power-law at low energies $\alpha_p<3$). However, regardless of the spectral effects, $\tau > 1$ implies that the photons are trapped inside the source {by scattering} unless the source {thickness} is $\Delta R < R/(2\Gamma^2 \tau^2)$. Therefore,  $\tau \gg 1$ leads to an unrealistically thin source, so also in case of a Comptonized spectrum the optical depth cannot be too large. In the limits that we derive below we set $\tau_{\rm min}=1$. If one is interested in the limit in case of a different value for $\tau_{\rm min}$ then the value of ${\cal L}$ in Table \ref{table summary2}  should be replaced by ${\cal L}/\tau_{\rm min}$. Given the weak dependence of the various limits on the value of ${\cal L}$, the effect on the result for any reasonable value of $\tau_{\rm min}$ is minor.

Taking $\tau_{\rm min}=1$, we  obtain first a  lower limit on the Lorentz factor for a given viewing angle.
Examination of Eq. \eqref{tau general} reveals a product of $\cal L$ that depends only on observational quantities and a term in square brackets that depends on observables via $f$ and on $\Gamma$ and $\theta$.
In the following we isolate the part that depends on the observed quantities in $f$ and we obtain an expression that depends just on $\Gamma $ and $\theta$ that we solve for each set of observables. 

This solution results in both a lower limit on $\Gamma$ and an upper limit of $\theta$. This is a result of the functional shape of the Doppler factor, which for $\Gamma\gg1$ and $\theta\ll1$ takes the form
\begin{align}
\delta_{\rm D}(\theta,\Gamma)\simeq\frac{2\Gamma}{1+(\Gamma\theta)^2}\propto\begin{cases}
\Gamma&; \theta\lesssim\Gamma^{-1},\\
\Gamma^{-1}\theta^{-2}&; \theta\gtrsim\Gamma^{-1}.
\end{cases}
   \label{doppler factor2}
\end{align}
For a given observing angle $\theta$, the Doppler factor has a maximum at {$\Gamma=1/\theta$.} It increases (decreases) with $\Gamma$ for on-axis (off-axis) observers. {Thus, for a given  $\delta_{\rm D}$ and a given  observing angle $\theta$ there are  either two solutions for $\Gamma$ 
 (an ``on-axis'' solution with $\theta < 1/ \Gamma $ and an ``off-axis'' solution with $ \theta > 1/ \Gamma$) or no solution if the observing angle is too large. }

This implies that compactness sets an upper limit on $\theta$, consider the dependence of the minimal optical depth, $\tau_{\rm min} $, on $\Gamma$
\begin{align}
\tau_{\rm min}\propto\frac{f}{\theta_\gamma^2\Gamma^2\delta_{\rm D}^4(\theta,\Gamma)}\propto\begin{cases}
f\Gamma^{-4}&; \theta\lesssim\Gamma^{-1},\\
f\Gamma^{2}\theta^{6}&; \theta\gtrsim\Gamma^{-1},
\end{cases}
\label{tau2}
\end{align}
where $\theta_\gamma={\rm max}\{1/\Gamma,\theta\}$.
The photon fraction, $f$, is a decreasing function of $\delta_{\rm D}$  (see Eqs. \ref{energy limit A}, \ref{energy limit B}, \ref{photon fraction PL}, \ref{photon fraction CPL}, and \ref{photon fraction limit C}) and hence for a fixed $\theta$ it is minimal at $\Gamma\simeq1/\theta$. Therefore, for any value of $\theta$ the optical depth $\tau_{\rm min}$ obtains a minimal value  for $\Gamma\simeq1/\theta$. If this value is larger than unity then there is no solution for this value of $\theta$. Now, setting $\Gamma=1/\theta$ shows that $\tau_{\rm min}$ increases monotonically with $\theta$. Therefore there is a maximal value of $\theta$ above  which $\tau_{\rm min}$ must be larger than unity and the compactness limit cannot be satisfied.
{The behavior of $\tau_{\rm min}$ (Eq. \ref{tau2}) is easily understood as follows. 
Observation of a given flux implies a photon and particle density in the source that is a rapidly increasing function of $\theta$ (because larger $\theta$ moves the observer out of the beam), and that above some value of $\theta$ the implied source density implies $\tau > 1$, forbidden by compactness.}

\begin{figure*}
\begin{center}
\includegraphics[width=85mm, angle=0]{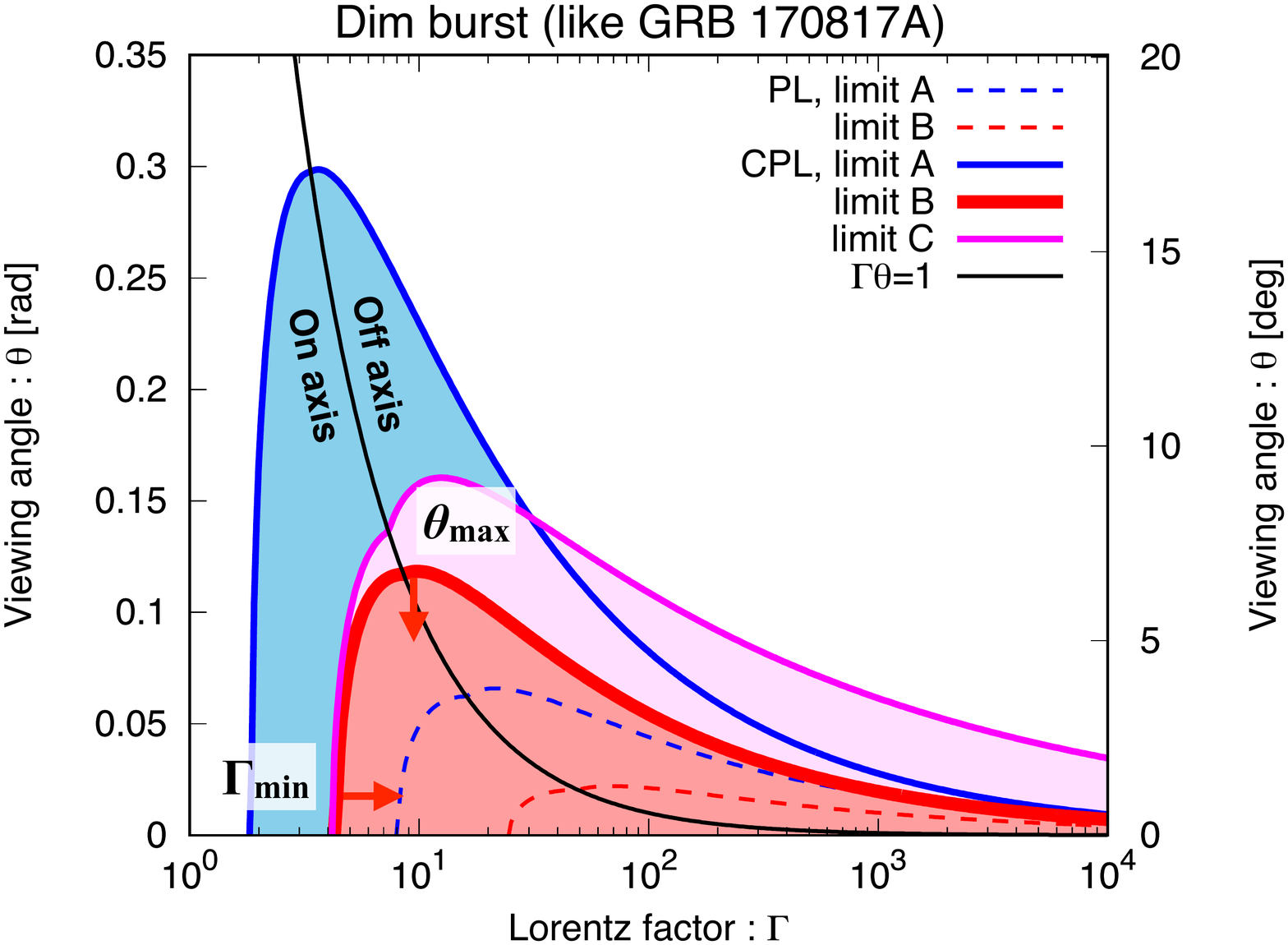}
\includegraphics[width=85mm, angle=0]{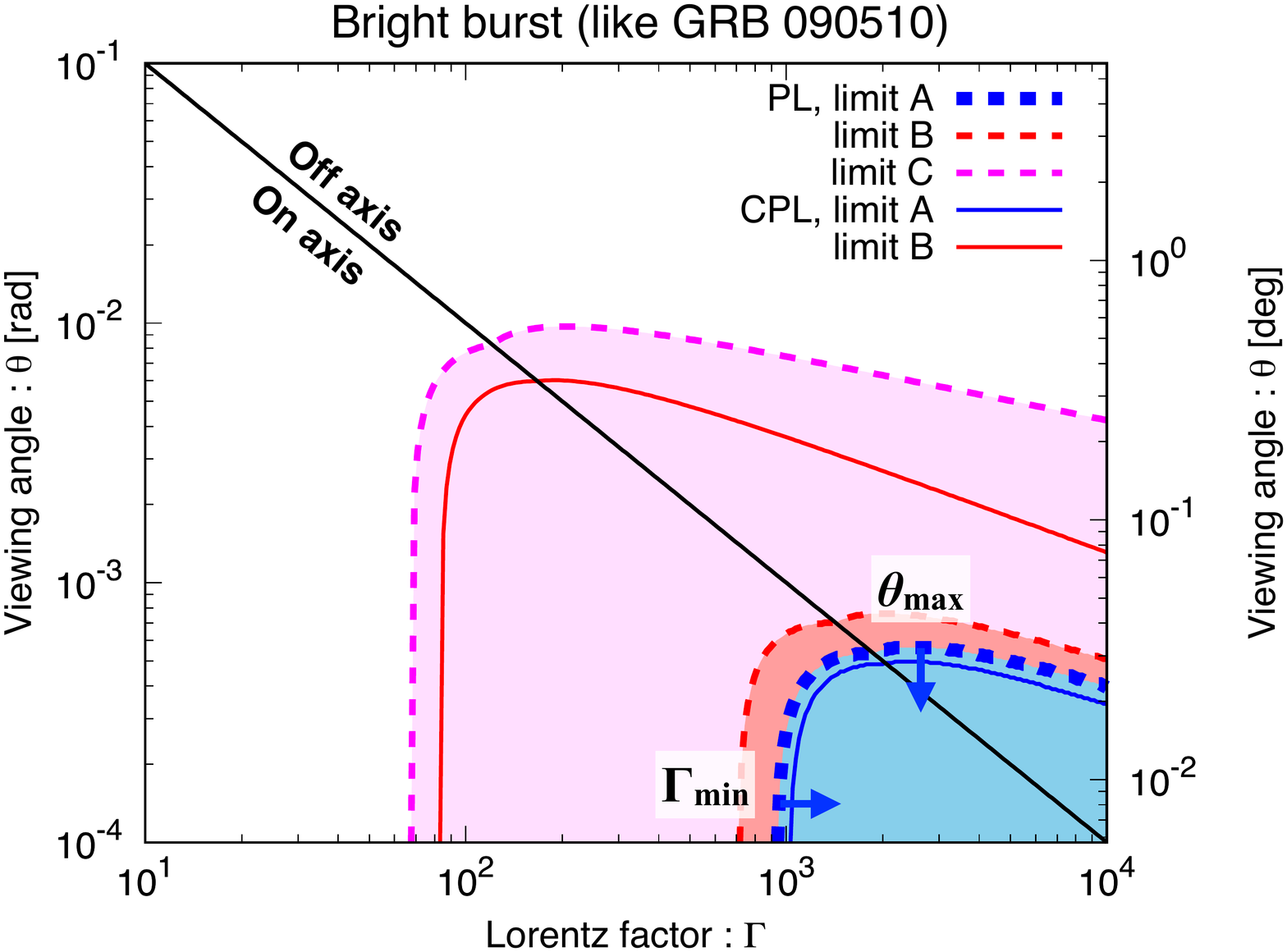}
\caption{The allowed Lorentz factor and viewing angle imposed by compactness for two types of GRBs (see Table \ref{table summary} for the specific parameters).
The dashed and solid curves show the limits for PL and CPL spectra, respectively, and the different colored curves denote different limits (see also Table \ref{table summary2}).
Colored regions show the allowed regions for the relevant spectral shape CPL (left) and PL (right) spectra. The other spectral limit is also shown for a comparison. 
The black curves ($\Gamma\theta=1$) divide the phase space to on and off-axis regions.
Limit C is independent of the spectral shape.
{\bf Left:} Parameters corresponding to the dim and soft sGRB 170817 A. 
Limit B (red curves) gives the most restrictive constraint.
{\bf Right:} Parameters corresponding to the bright and hard sGRB 090510.
Limit A (blue curves) gives the most severe limit.
In particular the viewing angle is constrained to be a very small value, $\theta\lesssim10^{-3}\,\rm rad$ (Note that the viewing angle is shown in a logarithmic scale in this figure).}
\label{fig sample}
\end{center}
\end{figure*}

This can be seen in Fig. \ref{fig sample} which depicts the allowed regions, for two sets of parameters (see Table \ref{table summary2}) resembling the observation of GRBs 170817A (left) and 090510 (right). The colored regions are values of $\Gamma$ and $\theta$ where $\tau$ can be smaller than unity and the thick lines mark the solution for $\tau_{\rm min}=1$. The allowed region is divided into on- and off-axis regions, where the minimal allowed Lorentz factor, $\Gamma_{\rm min}$ is  obtained for an on-axis observer. At $\theta=0$ we recover the  results of \cite{Lithwick&Sari2001}. Fig. \ref{fig sample} shows that for each model and each set of parameters there is a maximal value of $\theta$ above which there is no solution. Thus, while the compactness argument was brought up originally to put on a lower limit on the Lorentz factor, it turns out that at the same time it provides a corresponding limit on the maximal viewing angle   $\theta_{\rm max}$. 
This maximal angle  is obtained for an observer  located on the boundary between the ``on-axis'' and ``off-axis'' configurations, namely, for  $\theta \simeq 1/\Gamma$ and as we see later  $\theta \simeq 1/2\Gamma_{\rm min}$.

The compactness limits on $\Gamma$ and $\theta$ can be solved numerically for any set of parameters, however, it is illustrative to derive an analytic approximation. We therefore estimate the allowed regions of the different limits (A, B, and C) for a given set of parameters by approximating $\beta =1$ and $\theta_{\gamma}=1/\Gamma$ 
even for $\theta>1/\Gamma$, which does not change the result significantly.
With these approximations, the factor in square brackets in Eq. \eqref{tau general} depends only on the Doppler factor {for limits A and B.
We have an additional factor of $\Gamma$ for limit C.}
The optical depth is characterized by three dimensionless parameters: $\cal L$ (defined in    Eq. \eqref{L}), $\cal E$, and the spectral index $\alpha_p$ or $\beta_p$.
The dimensionless factor ${\cal E}$ is defined as:
\begin{align}
{\cal E}&\equiv\begin{cases}
\frac{\epsilon_{\rm p}\epsilon_{\rm max}(1+z)^2}{(\ME c^2)^2}&\text{; PL \& limit A,}\\
\frac{\epsilon_{\rm p}(1+z)}{\ME c^2}&\text{; PL \& limit B,}\\
\frac{\epsilon_{\rm p}\epsilon_{\rm max}(1+z)^2}{(\ME c^2)^2(\alpha_p+2)}&\text{; CPL \& limit A,}\\
\frac{\epsilon_{\rm p}(1+z)}{\ME c^2(\alpha_p+2)}&\text{; CPL \& limit B,}\\
\frac{\epsilon_{\rm p}(1+z)}{\MP c^2}&\text{; limit C.}
\end{cases}
   \label{E}
\end{align}
We have added here the redshift ($z$) dependence to the definition of $\cal E$. $\cal L$ is independent of the redshift.  There are three redshift effects \citep{Lithwick&Sari2001}:
(i) the distance $d$ in Eq. \eqref{photon number} should be replaced by the luminosity distance $d\to d_{\rm L}/(1+z)$.
(ii) the timescales in Eq. \eqref{tau general} should be modified to $\delta t\to \delta t/(1+z)$  to account for the cosmological time dilation.
As pointed out by \cite{Lithwick&Sari2001}, these two effects (i) and (ii) cancel out.   
(iii) the photon energy is also redshifted: $\epsilon\to\epsilon(1+z)$.
This last effect modifies the threshold energy in Eqs. \eqref{energy limit A} and \eqref{energy limit B}, and Eq. \eqref{photon fraction limit C}.
Therefore, as noted above, the redshift changes only the parameter $\cal E$. 

With these definitions, the  minimal optical depth (Eq. \ref{tau general}) is given by
\begin{align}
\tau_{\rm min}&=\begin{cases}
\frac{{\cal L}}{{\cal E}^{\beta_p+1}}{\delta_{\rm D}^{2(\beta_p-1)}}&\text{; PL \& limit A,}\\
\frac{{\cal L}}{{\cal E}^{\beta_p+1}}{\delta_{\rm D}^{\beta_p-3}}&\text{; PL \& limit B,}\\
\frac{{\cal L}}{{\cal E}^{\alpha_p}}{\delta_{\rm D}^{2(\alpha_p-2)}}\exp\Bigl(-\frac{\delta_{\rm D}^2}{{\cal E}}\Bigl)&\text{; CPL \& limit A,}\\
\frac{{\cal L}}{{\cal E}^{\alpha_p}}{\delta_{\rm D}^{\alpha_p-4}}\exp\Bigl(-\frac{\delta_{\rm D}}{{\cal E}}\Bigl)&\text{; CPL \& limit B,}\\
\frac{{\cal LE}}{\Gamma\delta_{\rm D}^5}&\text{; limit C} \ . 
\end{cases}
   \label{tau parameterized}
\end{align}

These equations can be rewritten as limits on the Doppler factor in term of the observed parameters for limits A and B.
In particular, for PL spectrum, we obtain by setting $\tau_{\rm \gamma,min}=1$:
\begin{align}
\delta_{\rm D,min}&=\begin{cases}
\big({\cal E}^{\beta_p+1}/{\cal L}\big)^{\frac{1}{2(\beta_p-1)}}&\text{; PL \& limit A,}\\
\big({\cal E}^{\beta_p+1}/{\cal L}\big)^{\frac{1}{\beta_p-3}}&\text{; PL \& limit B.}
\end{cases}
   \label{doppler factor3}
\end{align}

Once the Doppler factor, $\delta_{\rm D,min}$, is known, we have: 
\begin{align}
\theta(\Gamma)=\arccos\biggl[\frac{\Gamma\delta_{\rm D,min}-1}{\delta_{\rm D,min}(\Gamma^2-1)^{1/2}}\biggl]\,;\,\,\,\text{limits A \& B.}
   \label{theta contour}
\end{align}
The minimal Lorentz factor is obtained by setting $\theta=0$ as
\begin{align}
\Gamma_{\rm min}=\frac{\delta_{\rm D,min}^2+1}{2\delta_{\rm D,min}}  \quad \Big[ \simeq \delta_{\rm D,min} /2 \Big] \ .
   \label{minimal lorentz factor}
\end{align}
The maximal angle for a given $\delta_{\rm D,min}$ is
\begin{align}
\theta_{\rm max}=\arcsin\big(\delta_{\rm D,min}^{-1}\big)\quad  \Big[ \simeq 1/2\Gamma_{\rm min} \Big] \ ,
   \label{maximal angle}
\end{align}
and the Lorentz factor at the maximal angle is 
\begin{align} \Gamma(\theta_{\rm max})=1/\sin\theta_{\rm max} \quad \Big[ \simeq 2 \Gamma_{\rm min} \Big] .
\end{align}
Where the approximate values  in the square brackets are valid for $\delta_{\rm D,min} \gg 1$. 

For limit C, since an additional factor of $\Gamma$ appears in Eq. \eqref{tau parameterized}, we approximately derive the allowed region for on-axis and off-axis regions separately.
In the on-axis regime, using Eq. \eqref{doppler factor2} one obtains $\tau_{\rm min}=1$ for:
\begin{align}
\Gamma=\Gamma_{\rm min}\simeq\bigg(\frac{{\cal LE}}{32}\biggl)^{1/6}\,;\,\,\,\text{limit C} \ .
\end{align}
In the off-axis regime, $\tau_{\rm min}=1$ is obtained for an angle
\begin{align}
\theta(\Gamma)\simeq\bigg(\frac{{\cal LE}}{32}\biggl)^{-1/10}\Gamma^{-2/5} \ ,
\end{align}
which is a decreasing function of $\Gamma$. Since in this regime $\theta>1/\Gamma$ the maximal angle may be reasonably evaluated by Eq. \eqref{maximal angle}, $\theta_{\rm max}\simeq1/2\Gamma_{\rm min}$, as for limits A and B.

Tables \ref{table summary} and \ref{table summary2} summarize the results of this paper. 
In Table \ref{table summary2}, we list  the definitions of the parameters $\cal L$ and $\cal E$, that  are obtained by the observables in Table \ref{table summary}, and the expressions for the optical depth and the resulting limits on $\delta_{\rm D}$, $\Gamma$, and $\theta$ (if it is analytically derived) for each spectrum.

\begin{figure}
\begin{center}
\includegraphics[width=85mm, angle=0]{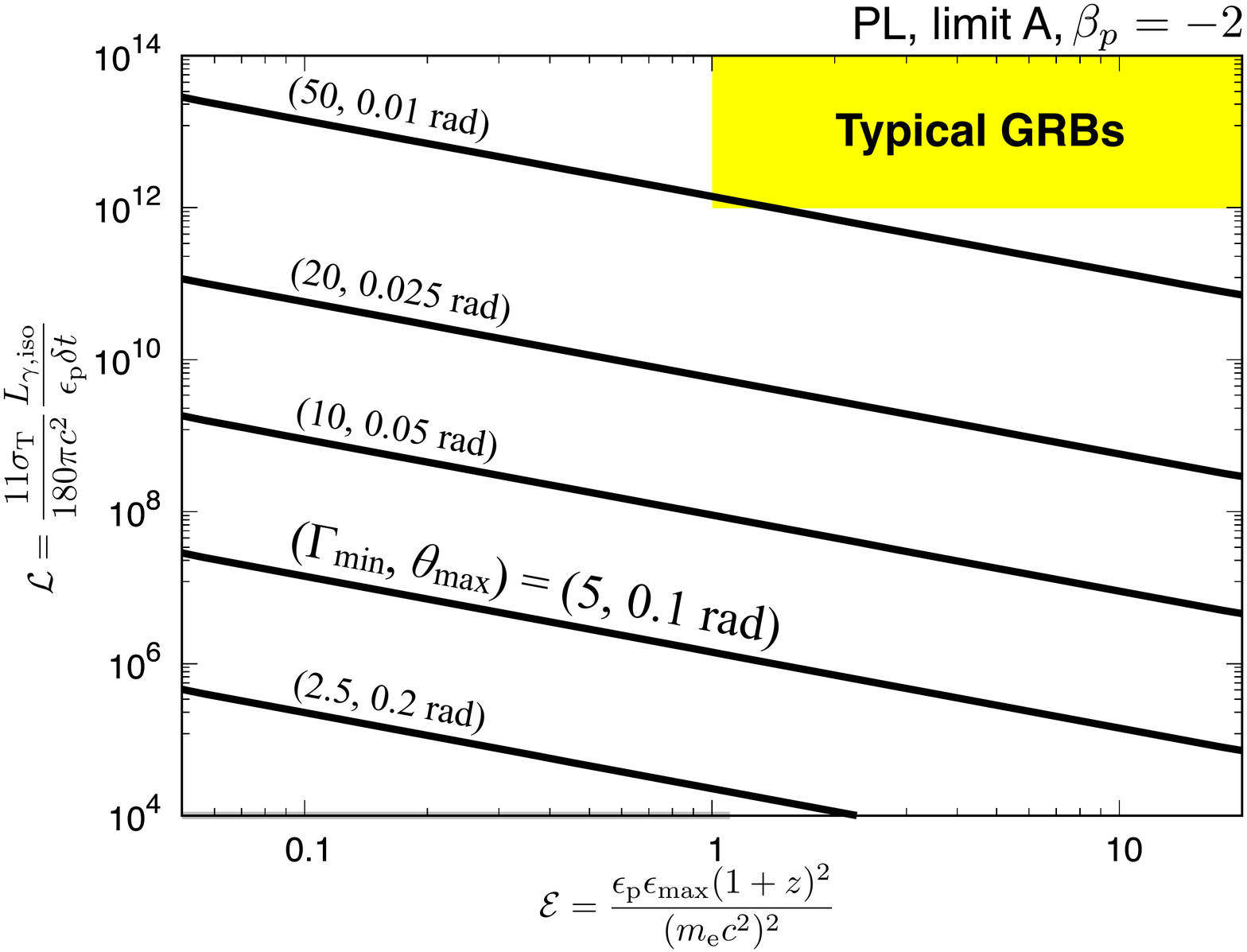}
\caption{Contours  lines of the maximal viewing angle, $\theta_{\rm max}$, and the minimal Lorentz factor, $\Gamma_{\rm min}$,  for a PL spectrum and limit A (the spectral index is fixed $\beta_p=-2$).
The yellow shaded region show the parameter space of normal GRBs with ${\cal L}\gtrsim10^{12}$ and demonstrates that typical GRBs are unlikely to be observed from an off-axis view.}
\label{fig theta_PL_A}
\end{center}
\end{figure}

Fig. \ref{fig theta_PL_A}, depicts the contours of a fixed  $\theta_{\rm max}$ and $\Gamma_{\rm min}$ as a function of the observables $\cal L$ and $\cal E$ for a PL spectrum in limit A, which is relevant for typical GRBs.
Due to the power-law dependence of $\tau_{\rm min} $ on $\cal L$ and $\cal E$, the contours are straight lines.
A small $\cal E$ implies  a large energy threshold for pair creation, which reduces the photon fraction and hence the compactness limits are relaxed leading to a lower values of $\Gamma_{\rm min}$ and larger values of $\theta_{\rm max}$.
The yellow shaded region in this figure corresponds to typical GRBs with ${\cal L}\gtrsim10^{12}$.
These GRBs have very small maximal viewing angles  $\theta_{\rm max}\lesssim10^{-2}\,\rm rad$. 
The chance to observe such a GRB off-axis (relative to observing it on-axis) is $\theta_{\rm max}/\theta_{\rm j}\ll 1 $ where 
$\theta_{\rm j}$ is the jet opening angle of the GRB. 
In other words, it is hardly possible that regular GRBs are viewed  off-axis (see also \S \ref{grb090510}).

\begin{figure}
\begin{center}
\includegraphics[width=85mm, angle=0]{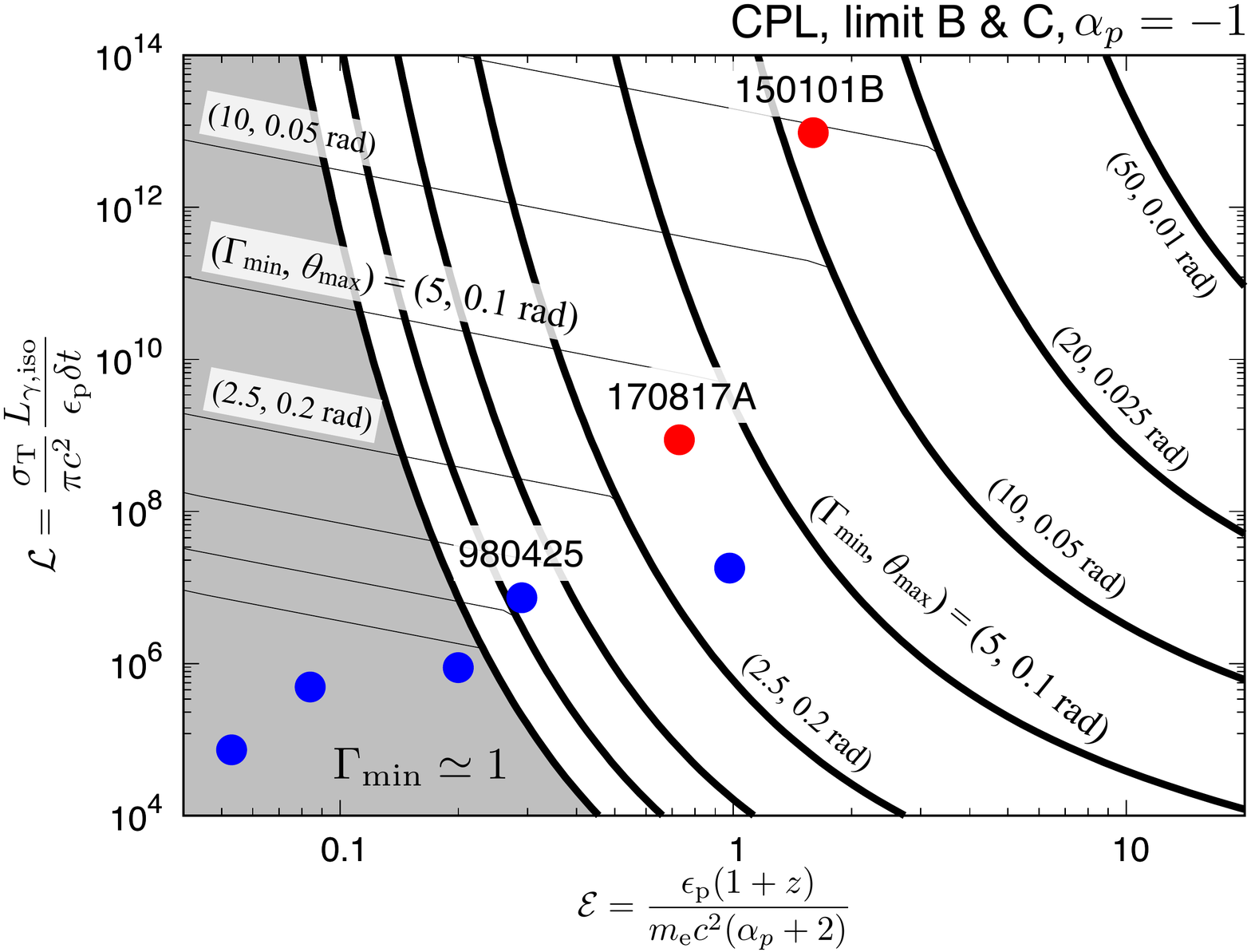}
\caption{The same as Fig. \ref{fig theta_PL_A} but for a CPL spectrum and limits B and C with $\alpha_p=-1$.
The thick curves denote the contours of limit B for fixed values of  $\theta_{\rm max}=$ 0.01, 0.025, 0.05, and $0.1$ to $0.5\,\rm rad$ with $0.1\,\rm rad$ steps and the corresponding $\Gamma_{\rm min}$  values. 
The contours for limit C are also shown with thin curves in the region where limit C is more constraining than limit B. The values of $\Gamma_{\rm min}$ and $\theta_{\rm max}$ of these contours are the same as those on the B contours in which they end.
The red and blue points show observed sGRB and \llgrbs, respectively.
The grey shaded region denotes a parameter region where a relativistic motion is not required. }
\label{fig theta_CPL}
\end{center}
\end{figure}

Fig. \ref{fig theta_CPL} depicts contour lines of $\Gamma_{\rm min}$ and $\theta_{\rm max}$ for a CPL spectrum in limits B  ($\alpha_p=-1$) and C.
These limits are relevant for low luminosity \gr transients (see \S \ref{sec:application}).
Since the CPL spectrum is exponential, the contours have no longer a simple form.
The contour's shape does not change much for a different value of $\alpha_p$ because the spectral index is absorbed into the definition of $\cal E$ and it disappears from the exponential.
Also shown are  $\Gamma_{\rm min}$ and $\theta_{\rm max}$ dictated by limit C.  This limit  becomes important for low luminosity bursts (e.g., left panel of Fig. \ref{fig sample}).
The contours for limit C are given by 
\begin{align}
{\cal L}\sim\frac{\MP\theta_{\rm max}^{-6}}{2\ME(\alpha_p+2){\cal E}}\simeq{9.2\times10^{8}}\,\biggl(\frac{\theta_{\rm max}}{0.1\,\rm rad}\biggl)^{-6} \frac{1}{(\alpha_p+2)\cal E}.
   \label{contour limit C}
\end{align}
Note that in this equation, the parameter $\cal E$ is defined for limit B (see Eq. \ref{E}) and overall the limit is independent of the spectral shape, as expected. 
Like in limit A, a smaller $\cal E$ values allow larger viewing angles.
However, if $\cal E$ is too small, limit C constrains the maximal angle more strictly.
We also find that for events with ${\cal L}\lesssim10^{7-8}$ and ${\cal E}\lesssim0.1$, the compactness considerations do not require a relativistic motion of the \gr sources.

\section{Application to $\gamma$-ray transients}
\label{sec:application}
We turn now to  apply the generalized compactness argument to  specific events (listed up in Table \ref{table events}) focusing on those events that  have been discussed in the context of off-axis emission models.

\begin{table*}
\begin{center}
\caption{Observables and the derived $\Gamma_{\rm min}$ of selected \gr transients. The most constraining limit is denoted by boldface. For GRB 170817A and 150101B, the first and second lines show the observables given by time-resolved spectral analysis and ones by time-integrated analysis. $^*$ Note that the analysis breaks down for $\Gamma_{\rm min}$ values approaching unity. }
\label{table events}
\begin{tabular}{l|rrrrrrr|ccc}
\hline\hline
\multirow{2}{*}{Event (Ref)}&\multicolumn{7}{c|}{Observables}&\multicolumn{3}{c}{Limit : $\Gamma_{\rm min}$($\sim1/2\theta_{\rm max}$)}\\
&$L_{\rm \gamma,iso}\,[\rm erg\, s^{-1}]$&$\delta t\,[\rm s]$&$\epsilon_{\rm p}\,[\rm keV]$&$\epsilon_{\rm max}\,[\rm keV]$&$\beta_p$ (PL)&$\alpha_p$ (CPL)&$z$&limit A&limit B&limit C\\
\hline
GRB 090510 (1) & $10^{53}$ &$0.01$ & $4000$ & $3\times 10^7$ & $-2$ &-&$0.9$& $\bf 910$ & $690$ & $66$\\
\hline
GRB 170817A (2,3,4) & $2\times10^{47}$ &$0.2$ & $520$ & $520$ & - & $-0.6$ &0.008& $1.7$ & $3.7$ & $\bf 4.2$\\
&&&$185$&& - &&& $1.3$ & $2.0$ & $\bf 4.2$\\
GRB 150101B (5) & $10^{51}$ & $0.01$ & $1300$ & $1300$ & - & $-0.8$&$0.1$ & $5.5$ & $17$ & $\bf 29$\\
& $4\times10^{50}$ & $0.02$ & $550$ &  & - & & & $3.6$ & $7.9$ & $\bf 21$\\
\hline
GRB 980425  (6,7,8) & $6\times10^{46}$ & $10$ & $70$ & $200$ & $-2.3$&-&0.009& $2.2$ & $\bf 3.7$ & $1.9^*$ \\
&&& $120$ & &-&$-1.2$&& $1^*$ & $1.5^*$ & $\bf 1.9^*$\\
GRB 031203 (9) & $10^{48}$ & $40$ &$>200$ & $400$ &-&$-1.6$&$0.1$& $1.5^*$ & $\bf 3.2$ & $2.3$\\
GRB 060218 (8,10) & $3\times10^{46}$ &$2000$ &$30$ & $100$ & - &$-0.9$& $0.03$ & $1^*$ & $1^*$ & $1^*$\\
GRB 100316D (11) & $3\times10^{46}$ &$300$ &$30$ &-& - & $-1.4$& $0.06$ & - & $1^*$ & $\bf 1.2^*$\\
GRB 171205A (12) & $10^{47}$ &$140$ &$120$& $1000$ & - &$-0.85$& $0.04$  & $1.2^*$ & $1.2^*$ & $\bf 1.4^*$\\
\hline
\multicolumn{11}{l}{{\bf Refs.} (1) \cite{Ackermann+2010}, (2) \cite{Goldstein+2017}, (3) \cite{Veres+2018}, (4) \cite{Abbott+2017c}, (5) \cite{Burns+2018},}\\
\multicolumn{11}{l}{(6) \cite{Galama+1998}, (7) \cite{Frontera+2000}, (8) \cite{Kaneko+2007}, (9) \cite{Sazonov+2004}, (10) \cite{Campana+2006},}\\
\multicolumn{11}{l}{(11) \cite{Starling+2011}, (12) \cite{D'Elia+2018}}\\
\end{tabular}
\end{center}
\end{table*}

\subsection{Typical GRBs}\label{grb090510}
We begin with a discussion of the compactness  constraint on ordinary bright GRBs.
These bursts have a large \gr luminosity $L_{\rm \gamma,iso}\gtrsim10^{51}\,\rm erg\,s^{-1}$ and a rapid time variability $\delta t
\sim0.01-0.1\,\rm s$. These give rise to a large ${\cal L}\gtrsim10^{12}$ (Since we consider limit A we add an additional factor $11/180$ to  Eq. \ref{L}).
The allowed parameter region for typical GRBs is shown in Fig. \ref{fig theta_PL_A}  and the specific limits on GRB 090510 \citep{Ackermann+2010} are shown on the right panel of Fig. \ref{fig sample}. Note that GRB 090510 is one of the brightest and hardest GRBs and hence its limits are tighter than those of a typical GRB.
The large value of $\cal L$ in typical GRBs implies that $\Gamma_{\rm min}$ and $\theta_{\rm max}$ are tightly limited as $\Gamma_{\rm min}\gtrsim10^2$ and $\theta_{\rm max}\lesssim10^{-2}\,\rm rad$.
The very small allowed viewing angle justifies the commonly adopted ``on-axis'' assumption and limits the chance that a typical GRB is viewed by an  off-axis observers.

For typical GRBs the commonly used limit A is the dominant one. Limit B is  
applicable for softer bursts and in particular for weak bursts with a CPL spectrum. 
Limit C is hardly important for typical bursts, but it is used, implicitly, in regular GRBs in the context of photospheric emission models.

\subsection{GRB170817A}\label{grb170817}
Next, we revisit the compactness of GRB 170817A \citep{Kasliwal+2017,Gottlieb+2018b,Matsumoto+2019}.
The very dim \gr luminosity motivated some authors to consider an off-axis emission model, where the \gr emission is produced from a relativistic jet core.
However, the constraint on the angular configuration from the compactness showed that the \gr emitting region should be closer to our line of sight than the jet core (\citealt{Matsumoto+2019}, see also \citealt{Ioka&Nakamura2019}).

\begin{figure}
\begin{center}
\includegraphics[width=80mm]{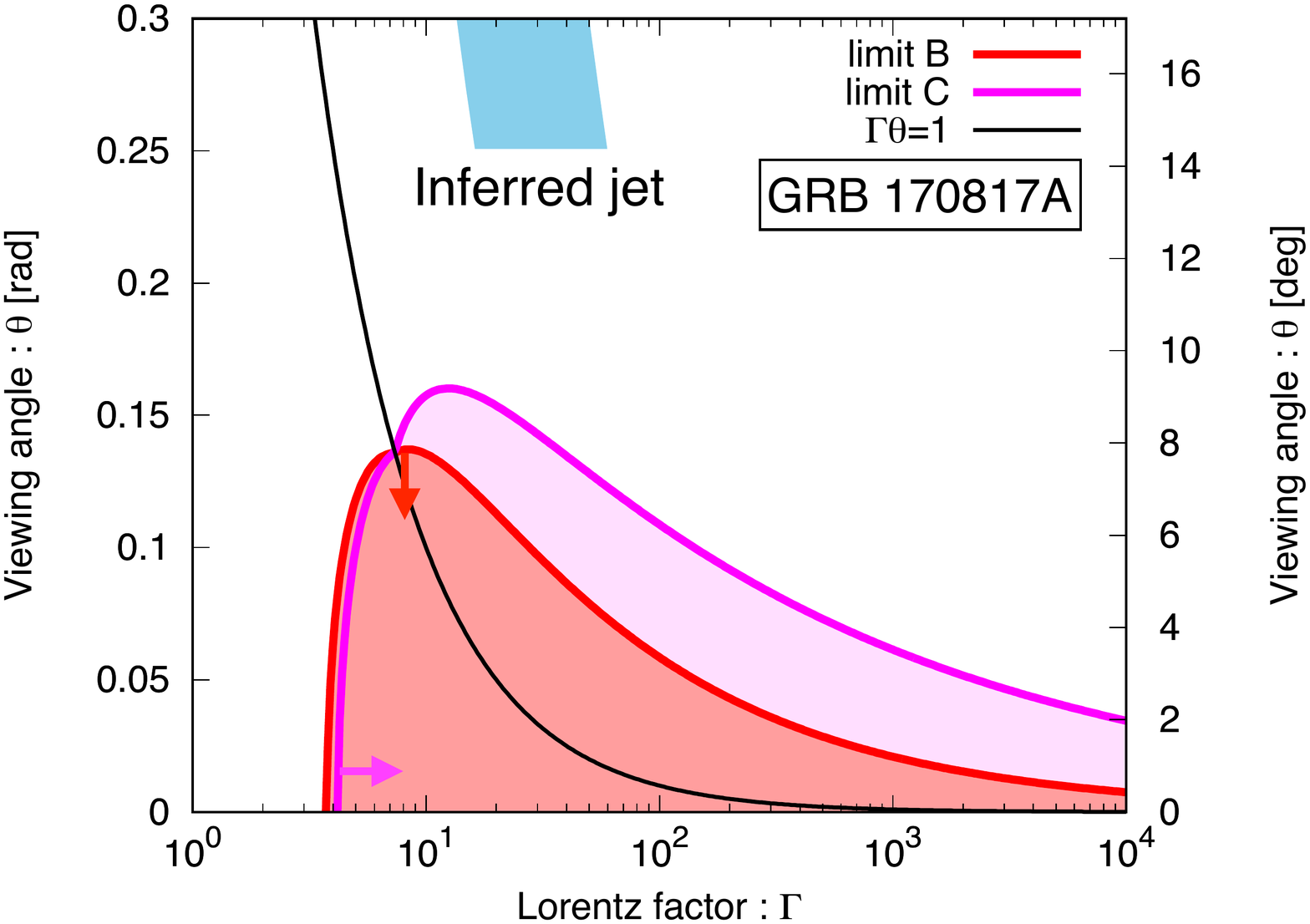}
\caption{The allowed region of the Lorentz factor and the viewing angle determined by the compactness argument for GRB 170817A for limits B (red) and C (magenta).
The parameters are set by $L_{\rm \gamma,iso}=2\times10^{47}\,\rm erg\,s^{-1}$, $\delta t
=0.2\,\rm s$, $\epsilon_{\rm p}=520\,\rm keV$, and $\alpha_p=-0.6$ (see Table \ref{table events}).
The blue shaded region depicts the required parameters for the emission to arise from the  jet that was inferred from the afterglow observations.}
\label{fig grb170817}
\end{center}
\end{figure}

Fig. \ref{fig grb170817} depicts the  allowed ($\Gamma, \theta_{\rm obs}$) region derived by imposing $\tau_{\rm min}<1$.
Since both limits B and C can be important for dim bursts (see below), we show the allowed region for both limits (red and magenta curves).
There is a maximal allowed viewing angle as we discussed in \S \ref{constraints}.
The Lorentz factor and viewing angle are constrained to be $\Gamma_{\rm min}=3.7\,(4.2)$ and $\theta_{\rm max}=0.13\,(0.17) \,\rm rad$ by limits B (C), respectively
We stress again that for limit B we assumes that the exponential photon spectrum extends to higher energies than the observed ones.\footnote{\S 4 of \cite{Matsumoto+2019} discusses a spectrum independent limit that pair production opacity  imposes on the source of emission in GW170817.} But there is no reason to expect a super-exponential spectrum at higher energies. On the other hand limit C is independent of the spectrum.
We also remark that the maximal angle given by limit B, $\theta_{\rm max}$ is  smaller by about a factor of 2 than that obtained by \cite{Matsumoto+2019} (see their Fig. 2).
{This is because of a different definition of the angle. Here $\theta$ is the measured from the center of the \gr emitting region while  in \cite{Matsumoto+2019} it is  the angular distance from the edge of the \gr emitting region.}
Moreover, \cite{Matsumoto+2019} have approximated the Doppler factor $\delta_{\rm D}(\theta,\,\Gamma)\sim2/(\Gamma\theta^2)$, and this approximation breaks down for  low Lorenz factor.

\subsection{GRB 150101B}\label{grb150101}
GRB 150101B is one of the nearest sGRBs \citep{Troja+2018b,Burns+2018}.
Its \gr {isotropic equivalent} energy $E_{\rm \gamma,iso}\sim10^{49}\,\rm erg$ is relatively small. It has a CPL spectrum with a  rather  large peak photon energy,   $\epsilon_{\rm p}\sim1300\,\rm keV$, and it is highly variable,   $\delta t
\sim0.01\,\rm s$ (see Table \ref{table events}). GRB 150101B also shows a bright optical and long-time X-ray afterglows. \
\cite{Troja+2018b} suggested that we have observed an off-axis emission  with angular distance of $\theta\sim10{\,\rm deg}\simeq0.18\,\rm rad$, the strong optical afterglow is a macronova signal and the long-time X-ray emission is the off-axis afterglow. 
Since GRB 150101B is not as dim as GRB 170817A,  compactness gives a tighter {constraint} on the maximal angle. 
We find that  $\theta_{\rm max}\lesssim0.05\,\rm rad$ (see Fig. \ref{fig theta_CPL}).
This constraint rules out the off-axis model, which requires a much larger viewing angle.
The minimal Lorenz factor is constrained to be $\Gamma_{\rm min}\simeq20$.

Several authors suggested that the origin of the \grs in GW170817 is the shock breakout of the cocoon \citep{Kasliwal+2017,Nakar+2018,Gottlieb+2018b,Bromberg+2018,Pozanenko+2018,Beloborodov+2018}  Interestingly, the observed \grs from GRB 150101B satisfies the closure relation for relativistic shock breakouts \citep{Nakar&Sari2012}. A breakout radius of $R \sim 10^{12}$ cm and a Lorentz factor of $\Gamma \sim 35$ are predicted to produce a $\delta t \sim 0.01$s pulse with $\epsilon_{\rm p} \sim 1300\,\rm keV$ and a total energy of {$E_{\gamma,\rm iso} \sim 10^{49}\,\rm erg$}. The breakout radius is similar to the one suggested for GW170817 \citep{Nakar+2018}. The higher Lorentz factor is expected if the breakout is driven by a cocoon and is observed at an angle that is closer to the jet axis than in GW170817, as implied by the afterglow of GRB 150101B. This supports a picture where both GRBs 170817A and 150101B are generated by shock breakouts, where the main difference is the angle with respect to the jet axis and thus the Lorentz factor of the shock.

\subsection{Low luminosity GRBs (\textit{ll}GRBs)}\label{grb980425}
\textit{ll}GRBs are a subclass of long GRBs (lGRBs) whose properties are very different from those of typical lGRBs, such as much lower luminosity $\sim10^{46-48}\,\rm erg\,s^{-1}$, lower peak photon energy $\lesssim100\,\rm keV$, a smooth light curve and a longer entire duration $\gtrsim100\,\rm s$ \citep{Kulkarni+1998,Campana+2006,Soderberg+2006b,Kaneko+2007,Bromberg+2011}.
Table \ref{table events} describes the observables of several \llgrbs.
In particular, the lower photon energy and lesser time variability suggest weak relativistic boosts.
However, both lGRBs and \llgrbs are associated with peculiar broad-line type Ic SNe, which suggests that they share the same progenitors \citep[e.g.,][]{Nakar2015}. 
A common hypothesis that \llgrbs are ordinary GRBs viewed  off-axis \citep{Nakamura1998,Eichler&Levinson1999,Woosley+1999,Ioka&Nakamura2001,Yamazaki+2003,Waxman2004} is problematic because their afterglow observations do not show the expected off-axis afterglow signature \citep[e.g.,][for GRB 060218]{Soderberg+2006b}.

\begin{figure}
\begin{center}
\includegraphics[width=80mm]{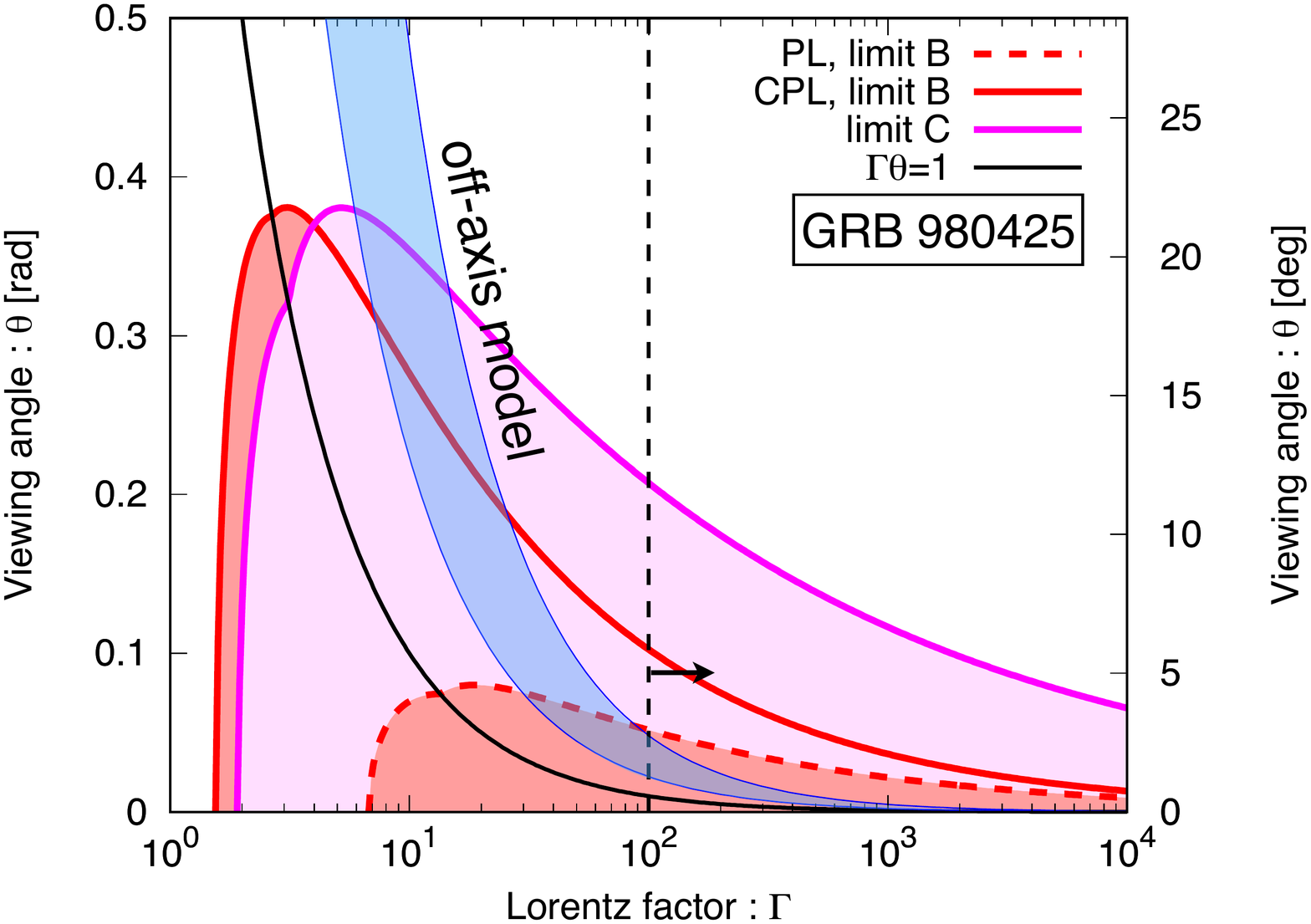}
\caption{The allowed region of the Lorentz factor and the viewing angle for \llgrb 980425.
The parameters are set by $L_{\rm \gamma,iso}=6\times10^{46}\,\rm erg\,s^{-1}$, $\delta t=10\,\rm s$, $\epsilon_{\rm p}=70\,\rm keV$ and $\beta_p=-2.3$ (for PL spectrum), and $120\,\rm keV$ and $\alpha_p=-1.2$ (for CPL spectrum) (see Table \ref{table events}).
In the off-axis model, the Lorentz factor and viewing angle are constrained within the blue shaded region.
The thin dashed vertical line marks the typical minimal Lorentz factor for regular GRBs. Thus if \llgrb 980425 was a regular lGRB viewed off-axis it should have been to the right of this line, allowing for a very small viewing angles, a possibility that is ruled out by radio afterglow observations \citep{Soderberg+2004}.  }
\label{fig grb980425}
\end{center}
\end{figure}

Fig. \ref{fig grb980425} shows the compactness limits B and C for \llgrb 980425.
As both PL and CPL spectra were fitted to this event (see Table \ref{table events}) we use both types of spectra.
The PL limits are quite constraining, but the CPL limits  as well as limit C that is independent of the spectrum show that a rather large region in the phase space is allowed with an observing angle of up to $\sim 0.4$ rad. 

Assuming that this burst was produced by an ordinary GRB with $E_{\rm \gamma,iso,obs}(\theta=0)\sim10^{50-52}\,\rm erg$ but viewed from off-axis \citep[see e.g.,][]{Ioka&Nakamura2001,Yamazaki+2003}
we estimate the allowed  parameter region for  the off-axis model  (shown as a blue shaded band in Fig. \ref{fig grb980425}). Indeed there is an overlap between the two regions, thus strictly speaking compactness cannot rule out the off-axis model. 
However the required region is constrained with a maximal viewing angle $\theta_{\rm max}\sim0.2-0.3\,\rm rad$. The {constraint is} more severe if we assume that the on-axis emission of \llgrb 980425 was a regular GRB with a typical Lorentz factor $\Gamma \gtrsim 10^2$. This would constrain the observing angle to be $\lesssim 0.06 $ rad, making the off-axis scenario extremely unlikely and already ruled out by radio afterglow observations \citep{Soderberg+2004}.

Like in GRB 980425 (see Fig. \ref{fig grb980425}), for other \llgrbs, compactness cannot exclude the off-axis scenario.
However, the compactness argument is still useful to constrain the allowed parameter region and becomes more powerful by combining afterglow observations.
As shown in Fig. \ref{fig theta_CPL}, most \llgrbs have larger maximal angle than that of GRB 980425 because of their small peak energy $\epsilon_{\rm p}$.
In particular, compactness does not necessarily require a relativistic outflow for GRBs 060218 and 100316D.

If we require ordinary lGRBs with $\Gamma\gtrsim10^2$ to produce the \llgrbs, the constraint on the viewing angle becomes more severe as shown for GRB 980425 in Fig. \ref{fig grb980425}. 
The maximal off-axis viewing angles for  \llgrbs 031203, 060218, 100326, and 171205, (assuming from  regular lGRB sources  with $\Gamma\gtrsim10^2$) are   $\theta_{\rm max}=0.064$, 0.28, 0.17, and 0.11 rad,  respectively. For GRB 031203, \cite{Ramirez-Ruiz+2005} gave $\theta_{\rm obs} \sim2\theta_{\rm j}\sim0.14\,\rm rad$. This is marginally excluded by compactness.
For GRB 060218, \cite{Soderberg+2006b} excluded the viewing angles $\theta\lesssim 1.1\,\rm rad$. Combining the compactness constraints, we can exclude the off-axis model for this event.   

While as stated above the compactness limit on its own cannot rule out the possibility that \llgrbs are regular lGRBs viewed off-axis, its implication are still far reaching. In particular, the very small viewing angle implied by this argument (see Figs. \ref{fig grb980425} and         \ref{fig theta_CPL})
and in particular the smaller angles implied with assuming typical lGRB sources with $\Gamma \gtrsim 10^2$ 
suggest that the rate of off-axis viewed lGRBs should be rather small as compared with the rate of regular lGRBs. On the other hand the observations imply that, while there are only few observed \llgrbs, their intrinsic rate is about ten times larger than the rate of lGRBs \citep[see e.g.][]{Soderberg+2006b}.

\section{Summary}\label{summary and discussion}
We extended here the compactness formalism, that was commonly used for on axis observers,  to an observer at a  general viewing angle.
We derived an expression for the optical depth of the \gr emitting region for three limits on the opacity \citep[see][]{Lithwick&Sari2001}:
Limit A is the commonly used one and it imposed that the highest energy photon can escape from the source.
The other limits are less known (see however \citealt{Lithwick&Sari2001}).
Limit B requires that the number of pairs produced is sufficiently small so that the system remains optically thin to Compton scattering.
In limit C, which is valid only for baryonic outflow, the electron accompanying the baryons should be optically thin for \grs.
We also considered two common spectral shapes (a single power-law and a  power-law with an exponential cut-off), and the analysis can be easily extended to other spectral shapes.  By imposing that the optical depth is smaller than unity, we obtain the allowed condition on the Lorentz factor $\Gamma$ and viewing angle $\theta$.
A given event can be characterized by two dimensionless parameters $\cal E$ and $\cal L$ (see Eqs. \ref{L} and \ref{E}), and the spectral index ($\alpha_p$ or $\beta_p$).  The  resulting limits  are summarized in Tables \ref{table summary} and \ref{table summary2}.

We find that the minimal Lorentz factor is obtained for an on-axis observer, $\theta=0$, this configuration is the same as the one discussed in  previous works \citep[e.g.,][]{Lithwick&Sari2001}. 
Interestingly, we find that there is a maximal allowed viewing angle $\theta_{\rm max}$ for each event.
This maximal viewing angle satisfies $\Gamma\theta_{\rm max}\simeq1$, thus it is on the boundary between being on-axis and off-axis. 
 
We applied the generalized compactness argument to the weak sGRBs 170817A and 150101B, and to several \llgrbs.
For GRB 170817A, we confirmed earlier results  \citep{Kasliwal+2017,Lazzati+2018,Kathirgamaraju+2018,Gottlieb+2018b,Bromberg+2018,Pozanenko+2018,Matsumoto+2019} that the observed relativistic jet core could not be the origin of the \grs in GRB 170817A.
For the  weak sGRB 150101B that was proposed to be an off-axis sGRB \citep{Troja+2018b,Burns+2018} we find a severe limit on the viewing angle, strongly disfavoring the off-axis option. 
For \llgrbs, we find that the off-axis emission scenario cannot be excluded by compactness argument alone. However, compactness arguments limit strongly the allowed parameter phase space for such models, which when combined with other constraints, make an off-axis origin highly unlikely.  Finally we find that regular GRBs (both long and short) can be viewed off axis only from a very small viewing angles. Thus the chance that we have observed a regular GRB off-axis is slim.

\section*{acknowledgments}
This research is supported by the CHE-ISF I-Core center for excellence in Astrophysics. 
TM is supported by JSPS Overseas Challenge Program for Young Researchers and 
 by Grant-in-Aid for JSPS Research Fellow 17J09895. 
 TP is supported by an advanced ERC grant TReX and by the Templeton foundation.
 EN is partially supported by an consolidator ERC grant (JetNS) and an ISF grant. 

\appendix 
\section{The light crossing time}\label{appendix a}

First we show that if  $\delta t_{\rm lc} >(\delta t_{\rm radial},\delta t_{\rm ang})$, i.e. the light crossing time is the longest, then the limits on the opacity are more constraining or equal to those obtained otherwise. Namely, that  
the least constraining limit on $\tau$ is obtained when $l^\prime \simeq \Delta R^\prime$. We will show that for the two limits of fully on-axis observer, $\theta=0$, and fully off axis observer, $\theta \gg 1/\Gamma$. Consider a photon released at a radius $R$ towards the observer from a point source that moves radially at a Lorentz factor $\Gamma$. By the time that the photon have reached $2R$ the point source have reached a radius $R(1 +\beta/\cos\theta)$, so the radial distance between the source and the photon is at this time is $d=R(1-\beta/\cos\theta)$, which for $\theta=0$ is $d \simeq R/2\Gamma^2$ and for $1/\Gamma\ll \theta \ll 1$ it is $d \simeq R \theta^2 /2$. Now, $l \simeq \min\{d,\Delta R\}$ and let's  define  $q >1$  such that $\Delta R=q d$, and $\delta t_{\rm lc}=q d/c$. This is compared to the radial time that for $dR=R$ is $\delta t_{\rm radial}=d /c$ (both on- and off-axis). Finally, $l^\prime/\Delta R^\prime=l/\Delta R=\min\{1,1/q\}$. With all these we can now 
compare $\tau$ between the case that the light crossing time dominates, i.e., $q>1$ and the case that the radial time dominates.  
From Eq. \eqref{tau_org} we see that for $q<1$ the optical depth is independent of $q$.  But, when $q>1$ the optical depth must be smaller by at least a factor of $q$. The reason is that when $q>1$ the limit $\delta t_{\rm radial}<\delta t_{\rm lc}$ implies that $R$ is smaller than $\Delta R$ by at least a factor of $q$. This implies that the least constraining limit on $\tau$ is obtained for $q<1$, namely the width of the emitting region is small enough so that the  light crossing time can be neglected.

\section{Angular timescale vs. radial timescale constrains}\label{appendix b}
We show first that in the off-axis case the limit on $\tau$ obtained under the assumption that $\delta t=\delta t_{\rm ang}$, is similar to the one obtained in case that $\delta t=\delta t_{\rm rad}$ and $dR=R$.
In off-axis case with $1/\Gamma \ll \theta \ll 1$, the angular time and radial time are approximately 
\begin{align}
\delta t_{\rm ang} = \frac{2R}{c}\sin(\theta)\sin\biggl(\frac{\theta_\gamma}{2}\biggl) \simeq \frac{R\theta\theta_\gamma}{c},
\end{align}
and 
\begin{align}
\delta t_{\rm radial}\simeq \frac{dR\theta^2}{c},
\end{align}
respectively. Note that for brevity, we ignore  here numerical factors of order unity. We focus on the factor $1/(\theta_\gamma R)^2$ in the optical depth (Eq. \ref{tau}). All  other factors are identical for both cases. When $\delta t=\delta t_{\rm ang}$, this factor becomes $1/(\theta_\gamma R)^2\simeq(\theta/c\delta t)^2$. On the other hand, when $\delta t=\delta t_{\rm radial}$ with $dR=R$, we obtain the same expression: 
\begin{align}
\frac{1}{(\theta_\gamma R)^2}\simeq\biggl(\frac{\theta^2}{\theta_\gamma c \delta t}\biggl)^2=\biggl(\frac{\theta}{c\delta t}\biggl)^2,
\end{align}
where we used $\theta_\gamma=\max (\theta, 1/\Gamma) = \theta$, as the larger value of $\theta_\gamma$ minimizes $\tau$. 

The limit on $\tau$ becomes similar in both cases for an on-axis observer ($\theta\ll1/\Gamma$), unless there is an anomalously narrow source with $\theta_\gamma \ll 1/\Gamma$. The timescales are given by 
\begin{align}
\delta t_{\rm ang}=\frac{2R}{c}\biggl\{1-\cos\biggl[\frac{1}{2}\biggl(\theta+\frac{\theta_\gamma}{2}\biggl)\biggl]\biggl\}\simeq \frac{R\theta_\gamma^2}{c},
\end{align} and 
\begin{align}\delta t_{\rm radial}\simeq \frac{dR}{c\Gamma^2}.
\end{align}
While the assumption of $\delta t=\delta t_{\rm ang}$ gives the factor 
\begin{align}
\frac{1}{(\theta_\gamma R)^2}\simeq\biggl(\frac{\theta_\gamma}{c\delta t}\biggl)^2,
\end{align} the case $\delta t=\delta t_{\rm radial}$ with $dR=R$ gives 
\begin{align}
\frac{1}{(R\theta_\gamma)^2}\simeq\biggl(\frac{1}{\theta_\gamma\Gamma^2c\delta t}\biggl)^2.
\end{align}
The factor $1/(\theta_\gamma R)^2$ becomes the same expression for both cases for an emitting region with a reasonable angular extent $\theta_\gamma\sim1/\Gamma$ .

\bibliographystyle{mnras}
\bibliography{reference_matsumoto}

\bsp	
\label{lastpage}
\end{document}